\documentstyle[epsfig,12pt]{article}

\begin{document}
                Submitted to "Il Nuovo Cimento C"
\begin{center}

{\bf

\title "GYRO-MAGNETIC RELATIONS AND  MASSES OF STARS}

\author "B.V.Vasiliev

Institute in Physical and Technical Problems,Dubna,Russia,141980

$e-mail: vasiliev@main1.jinr.ru$

\end{center}

\begin{abstract}
The calculations in Thomas-Fermi approximation show that in a
gravitational field each cell of ultra dense matter inside
celestial bodies obtains a very small positive electric charge. A
celestial body is electrically neutral as a whole, because the
negative electric charge exists on its surface. On the order of
magnitude the positive volume charge is very small ($10^{-18}e$
only). But it is sufficient to explain the occurrence of magnetic
fields of celestial bodies and the existence of a discrete
spectrum of steady-state values of masses of  stars and pulsars.

\end{abstract}


\clearpage

\section{}

We cannot measure magnetic fields of the majority of stars, which
are distant far from us. Therefore, the existence of magnetic
fields for the majority of stars can be considered only
hypothetically. However, the magnetic field of the Sun is known
over than a hundred years, and in the last decades the astronomers
managed to measure magnetic fields for a number of stars
(so-called $A_{p}$-stars) \cite{1} and some pulsars \cite{2}. It
is interesting to construct a model describing the generation of
magnetic fields by stars and to compare it with the data of the
astronomers. The mechanism examined below  is based on the
gravity-induced electric polarization of matter. It is capable to
explain also the generation of magnetic fields by planets
\cite{3}, however, in the case of stars, this mechanism works in
the purest manner.

\section{}
The action of gravity on metals has often been a topic of
discussion before \cite{4}-\cite{9}. The basic result of these
researches is reduced to the statement that gravity induces inside
a metal an electric field with an intensity

\begin{equation}
\vec{E}\simeq\frac{m_{i}\vec{g}}{e},\label{1010}
\end{equation}

where $m_{i}$ is the mass of an ion,

$\vec{g}$ is gravity acceleration,

$e$ is the electron charge.

This field is so small that it is not possible to measure it
experimentally. It is a direct consequence of the presence of an
ion lattice in a metal. This lattice is deformed by gravity and
then the electron gas adapts its density to this deformation. The
resulting field becomes very small.

Under superhigh pressure, all substances transform into ultradense
matter usually named nuclear-electron plasma \cite {10}. It occurs
when external pressure enhances the density of matter several
times \cite{10,11}. Such values of pressure exist inside celestial
bodies.

In nuclear-electron plasma the electrons form the degenerated
Fermi gas. At the same time, the positively charged ions form
inside plasma a dense packing lattice \cite{12},\cite{13}. As
usually accepted, this lattice may be replaced by a lattice of
spherical cells of the same volume. The radius $r_{s}$ of such a
spherical cell in plasma of the mass density $\gamma$ is given by

\begin{equation}
\frac{4\pi}{3}r_{s}^{3}=\biggl(\frac{\gamma}{m_{i}}\biggr)^{-1}=
\frac{Z}{n},\label{1020}
\end{equation}

where Z is the charge of the nucleus, $m_{i}=Am_{p}$ is the mass
of the nucleus, A is the atomic number of the nucleus, $m_{p}$ is
the mass of a proton, and n is the electron number density

\begin{equation}
n=\frac{3Z}{4\pi{r_{s}^{3}}}.\label{1030}
\end{equation}

The equilibrium condition in matter is described by the constancy
of its electrochemical potential \cite{10}. In plasma, the direct
interaction between nuclei is absent, therefore the equilibrium in
a nuclear subsystem of plasma (at $T=0$) looks like

\begin {equation}
\mu_{i}=m_{i}\psi+Ze\varphi=const.\label{1040}
\end {equation}

Here $\varphi$ is the potential of an electric field and $\psi$ is
the potential of a gravitational field.

The direct action of gravitation on electrons can be neglected.
Therefore, the equilibrium condition in the electron gas is

\begin {equation}
\mu_{e}=\frac{p_{F}^{2}}{2m_{e}}-(e-\delta{q})\varphi=const,\label{1050}
\end {equation}

where $m_{e}$ is the mass of an electron and $p _ {F} $  is the
Fermi momentum.

By introducing the charge $\delta {q}$, we take into account that
the charge of the electron cloud inside a cell can differ from
$e$. A small number of electrons can stay on the surface of a
plasma body where the electric potential is absent. It results
that the charge in a cell, subjected to the action of the electric
potential, is effectively decreased on a small value $ \delta
{q}$.

The electric polarization in plasma is a result of changing in
density of both nuclear and electron gas subsystems. The
electrostatic potential of the arising field is determined by the
Gauss' law

\begin{equation}
\nabla^{2}\varphi=\frac{1}{r^{2}}\frac{d}{dr}\biggl[r^{2}\frac{d}
{dr}\varphi\biggr]= -4\pi\biggl[Ze\delta(r)-en\biggr],\label{1060}
\end{equation}

where the position of nuclei is described by the function
$\delta(r)$.

According to the Thomas - Fermi method, $n$ is approximated by

\begin{equation}
n=\frac{8\pi}{3h^{3}}p^{3}_{F}.\label{1070}
\end{equation}

With this substitution, Eq.({\ref{1060}}) is converted into a
nonlinear differential equation for $\varphi$, which for $r>0$ is
given by

\begin{equation}
\frac{1}{r^{2}}\frac{d}{dr}\left(r^{2}\frac{d}{dr}\varphi(r)\right)=
4\pi\left[\frac{8\pi}{3h^{3}}\right] \left[2m_{e}(\mu_{e}+(e-
\delta{q})\varphi)\right]^{3/2}.\label{1080}
\end{equation}

It can be simplified by introducing the following variables
\cite{10}:

\begin{equation}
\mu_{e}+(e-\delta{q})\varphi=Ze^{2}{\frac{u}{r}}\label{1090}
\end{equation}

and $r=ax$,

where

$a=\{\frac{9\pi^{2}}{128Z}\}^{1/3}a_{0}$

with $ a_{0}=\frac{\hbar^{2}}{m_{e}e^{2}}=$ Bohr radius.

With the account of Eq.({\ref{1040}})

\begin{equation}
Ze^{2}{\frac{u}{r}}= const
-\frac{m_{i}\psi}{Z}-\delta{q}\varphi.\label{1092}
\end{equation}

Then Eq.({\ref{1080}}) gives

\begin{equation}
\frac{d^{2}u}{dx^{2}}=\frac{u^{3/2}}{x^{1/2}}.\label{1100}
\end{equation}

In terms of u and x, the electron density within a cell is given
by \cite{10}

\begin{equation}
n_{TF}=\frac{8\pi}{3h^{3}}p^{3}_{F}=
\frac{32Z^{2}}{9\pi^{3}a^{3}_{0}}
\biggl(\frac{u}{x}\biggr)^{3/2}.\label{1110}
\end{equation}

Under the influence of gravity the charge of the electron gas in a
cell becomes equal to

\begin{equation}
Q_{e}=4\pi{e}\int^{r_{s}}_{0}n(r)r^{2}dr=\frac{8\pi{e}}{3h^{3}}
\biggl[2m_{e}\frac{Ze^{2}}{a}\biggr]^{3/2}4\pi{a}^{3}
\int^{x_{s}}_{0}x^{2}dx\biggl[\frac{u}{x}\biggr]^{3/2}.\label{1140}
\end{equation}

Using Eq.({\ref{1100}}), we obtain

\begin{equation}
Q_{e}=Ze\int^{x_{s}}_{0}xdx\frac{d^{2}u}{dx^{2}}=
Ze\int^{x_{s}}_{0}dx\frac{d}{dx}\biggl[x\frac{du}{dx}-u\biggr]=
Ze\biggl[x_{s}\frac{du}{dx}\bigg|_{x_{s}}-u(x_{s})+u(0)\biggr].
\label{1150}
\end{equation}

At $ r\rightarrow0 $ the electric potential is due to the nucleus
alone $ \varphi{(r)} \rightarrow\frac {Ze} {r} $. It means that $
u(0)\rightarrow1 $ and each cell of plasma obtains a small charge

\begin{equation}
\delta{q}=Ze\biggl[{x_{s}}\frac{du}{dx}\bigg|_{x_{s}}-u(x_{s})\biggr]=
Ze{x_{s}}^2\biggl[\frac{d}{dx}\biggl(\frac{u}{x}\biggr)\biggr]_{x_{s}}.
\label{1160}
\end{equation}

For a cell placed in a point $R$ inside a star

\begin{equation}
\delta{q}=Zer_{s}^2\biggl[\frac{d}{dR}
\biggl(\frac{u}{r}\biggr)\biggr]\biggl[\frac{dR}{dr_{s}}\biggr].
\label{1170}
\end{equation}

Considering that gravity acceleration $\vec{g}=-\frac{d\psi}{dR}$
and the electric field intensity $\vec{E}=-\frac{d\varphi}{dR}$

\begin{equation}
\frac{dr_{s}}{dR}=\frac{r_{s}^2}{e}\biggl[\frac{\frac{m_{i}}{Z}\vec{g}
+\delta{q}\vec{E}}{\delta{q}}\biggr].\label{1180}
\end{equation}

This equation has the following solution

\begin{equation}
\frac{dr_{s}}{dR}=0\label{1190}
\end{equation}

and

\begin{equation}
\frac{m_{i}}{Z}\vec{g}+\delta{q}\vec{E}=0.\label{1200}
\end{equation}

In plasma, the equilibrium value of the electric field on nuclei
according to Eq.({\ref{1040}}) is determined by Eq.({\ref{1010}})
as well as in a metal. But there is one more additional effect in
plasma. Simultaneously with the supporting of nuclei in
equilibrium, each cell obtains an extremely small positive
electric charge.

As $div{\vec{g}}=-4\pi{G}{n}m_{i}$ and
$div{\vec{E}}=4\pi{n}\delta{q}$, the gravity-induced electric
charge in a cell

\begin{equation}
\delta{q}=\sqrt{G}\frac{m_{i}}{Z}\simeq{10^{-18}e},\label{1210}
\end{equation}

where $G$ is the gravity constant.

However, because the sizes of bodies may be very large, the
electric field intensity may be very large as well

\begin{equation}
\overrightarrow{E}=\frac{\overrightarrow{g}}{\sqrt{G}}.\label{2050}
\end{equation}

In accordance with Eqs.({\ref{1190}},{\ref{1200}}), the action of
gravity on matter is compensated by the electric force and the
gradient of pressure is absent.

Thus, a celestial body is electrically neutral as a whole, because
the positive volume charge is concentrated inside the charged core
and the negative electric charge exists on its surface and so one
can infer gravity-induced electric polarization  of a body.

\section{}

At the surface of the core, the electric field intensity reduces
to zero. The jump in electric field intensity is accompanied at
the surface of the core by the pressure jump $\Delta p(R_{N})$. It
leads to the redistribution of the matter density inside a star.
In a celestial body consisting of matter with an atomic structure,
density and pressure grow monotonously with depth. In a celestial
body consisting of electron-nuclear plasma, the pressure gradient
inside the polarized core is absent and the matter density is
constant. Pressure affecting the matter inside this body is equal
to the pressure jump on the surface of the core

\begin{equation}
p=\Delta p(R_{N})=\frac{E(R_{N})^{2}}{8\pi }=\frac{2\pi}{9}G\gamma
^{2}R_{N}^{2},  \label{210}
\end{equation}

where R$_{N}$ is the radius of the core.

One can say that this pressure jump is due to the existence of the
polarization jump or, which is the same, the existence of the
bound surface charge formed by an electron pushed out from the
core and making the total charge of the celestial body equal to
zero.

Because the electron subsystem of plasma inside a star is the
relativistic Fermi gas, we can write its  equation of state
\cite{10}

\begin{equation}
p=\frac{(3\pi^{2})^{1/3}}{4}\frac{{\hbar}c\gamma^{4/3}}{{m_{p}}^
{4/3}\beta^{4/3}}\label{220}
\end{equation}

where $\beta\cdot{m_{p}}$ is the mass of the matter related to one
electron of the Fermi gas system, and

$m_{p}$ is the proton mass.

Because of the electroneutrality, one proton should be related to
electron of the Fermi gas of plasma. The existence of one neutron
per proton is characteristic for a substance consisting of light
nuclei. The quantity of neutrons grows approximately to 1.8 per
proton for the  heavy nuclei substance. Therefore, it is necessary
to expect that inside stars

\begin{equation}
2<\beta<2.8 .\label{222}
\end{equation}

As pressure inside a star is known (Eq.({\ref{210}})), from
Eq.({\ref{220}}) it is possible to determine a steady-state value
of mass of a star

\begin{equation}
M_{\star}=\zeta{A_{\star}^{3/2}}\frac{m_{p}}{\beta^2}.\label{230}
\end{equation}

This mass is expressed by dimensionless constants only

\begin{equation}
A_{\star}=\biggl(\frac{\hbar{c}}{G{m_{p}}^2}\biggr)=1.54\cdot{10^{38}}\label{240}
\end{equation}

$\zeta=(1.5^5\pi)^{1/2}\simeq{5}$,

and the slowly varying parameter $\beta$ (Eq.({\ref{222}})).

The masses of stars can be measured with a considerable accuracy,
if these stars compose a binary system. There are almost 200
double stars whose masses are known with the required accuracy
\cite{15}. Among these stars there are giants, white dwarfs, and
stars of the main sequence. Their averaged mass is described by
the equality

\begin{equation}
\langle M_{\star}\rangle =\left( 1.36\pm 0.05\right) M_{\odot },
\label{250}
\end{equation}

where $M_{\odot }$ is the mass of the Sun.

The center of this distribution (Fig.1) corresponds to
Eq.({\ref{230}}) at $\beta \simeq 2.6$.

\begin{figure}
\begin{center}
\includegraphics[5cm,14cm][17cm,2cm]{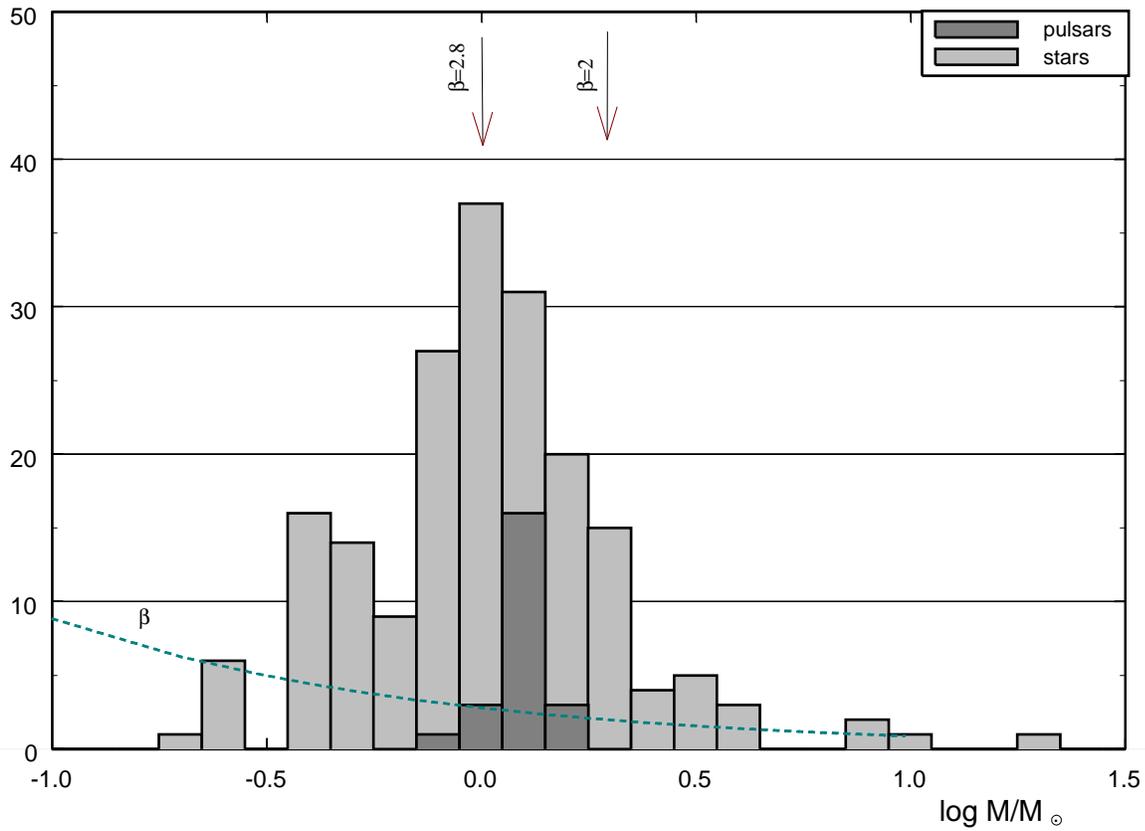}
\vspace{11cm}
\caption{Mass distributions of stars and pulsars
from the binary systems \cite{15},\cite{17}. The curve shows
Eq.({\ref{230}}).} \label{fig1}
\end{center}
\end{figure}

It is interesting to note that the "biography" of such a star
appears much poorer than in the Chandrasecar model.

Temperature does not influence the parameters of relativistic
plasma. Therefore, a star with a mass close to the steady-state
value (Eq.({\ref{230}})) is in a stable equilibrium not depending
on temperature. It should not collapse with a temperature
decreasing. The instability of a star can arise with burning out
of light nuclei - deuterium and helium - and with a related
increasing of $\beta$. This growth leads to the reduction of a
steady-state value of mass (Eq.({\ref{230}})) and, probably, to
the distraction of stars with greater masses.

\section{}

As the density of matter inside a relativistic star is constant,
it is possible to assume that it equals the mean density of the
Sun and to estimate a star radius

\begin{equation}
R\simeq\biggl(\frac{M_{\star}}{\frac{4\pi}{3}
{\gamma_{\odot}}}\biggr)^{1/3},\label{310}
\end{equation}

where $\gamma_{\odot}$ is the mean density of the Sun.

It allows one to calculate the momentum of a star as the momentum
of a sphere with a constant density

\begin{equation}
I=\frac{2}{5}M_{\star}{R^2}\label{320}
\end{equation}

and at a known frequency of rotation $\Omega$ to calculate its
angular momentum

\begin{equation}
L=\frac{2}{5}M_{\star}\Omega{R^2}.\label{330}
\end{equation}

In this model the magnetic moment of a star is created by the
rotation of a star as a whole. Thus, it is composed of two parts.
One is the magnetic moment of the layer of electrons placed on the
external surface of a star

\begin{equation}
\mu_{-}=-\frac{1}{3}\biggl({\frac{4\pi}{3}\rho{R^3}}\biggr)\Omega{R^2}.\label{340}
\end{equation}

The second component of the magnetic moment is created by the
positively charged core

\begin{equation}
\mu_{+}=\frac{1}{5}\biggl({\frac{4\pi}{3}\rho{R^3}}\biggr)\Omega{R^2}.\label{350}
\end{equation}

The summary moment is
\begin{figure}
\begin{center}
\includegraphics[5cm,14cm][17cm,2cm]{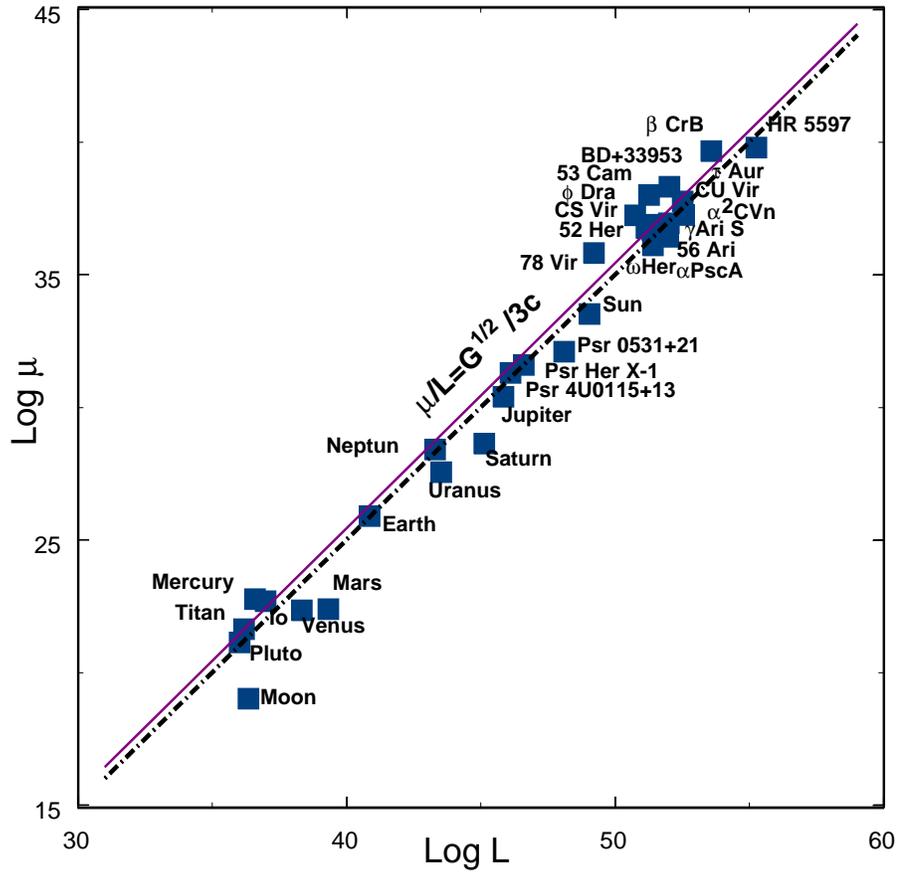}
\vspace{11cm} \caption{The observed values of the magnetic moments
of celestial bodies vs. their angular momenta. On the ordinate,
the logarithm of the magnetic moment over $Gs\cdot{cm^3}$ is
plotted; on the abscissa the logarithm of the angular momentum
over $erg\cdot{s}$ is shown. The solid line illustrates
Eq.({\ref{370}}). The dash-dotted line is the fitting of the
observed values.} \label{fig2}
\end{center}
\end{figure}

\begin{equation}
\mu_{\Sigma}=-\frac{2}{15}\biggl({\frac{4\pi}{3}\rho{R^3}}\biggr)
\Omega{R^2}.\label{360}
\end{equation}

It is remarkable that the gyromagnetic relation of a star, i.e.,
the relation of its magnetic moment to the angular momentum, is
expressed  through world constants only

\begin{equation}
\vartheta=\frac{\mu_{\Sigma}}{L}=\frac{\sqrt{G}}{3c}.\label{370}
\end{equation}

The measurements permit us to define the frequency of rotation and
magnetic fields for a number of stars \cite {1}. It appears enough
to check up the considered theory, since masses of stars and their
momenta are determined inside the theory (Eq.({\ref{250}}) and
(Eq.({\ref{330}}))). The magnetic moments as functions of their
angular momenta for all celestial objects (for which they are
known today) are shown in Fig.2. The data for planets are taken
from \cite{16}, the data for stars are taken from \cite{1}, and
for pulsars - from \cite{2}. As it can be seen from this figure
with the logarithmic accuracy, all celestial bodies - stars,
planets, and pulsars - really have the gyromagnetic ratio close to
the universal value (Eq.({\ref{370}})). Only the data for the Moon
fall out, because its size is too small to create an electrically
polarized core.

\section{}

Apparently, the considered theory is quite true for pulsars which
consist, as it is supposed, from the neutron substance with an
addition of electrons and protons \cite{10}. As this substance is
a relativistic one, there is a fair definition of a steady-state
value of mass Eq.({\ref{230}}). The astronomers measured masses of
16 radio-pulsars and 7 x-ray pulsars included in a double system
\cite {17}. According to this data, the distribution of masses of
pulsars is

\begin{equation}
\langle M_{pulsar}\rangle=(1.38\pm{0.03})M\odot.\label{380}
\end{equation}

The center of this distribution corresponds to Eq.({\ref{230}}) at
$\beta \simeq 2.6$.

\begin{figure}
\begin{center}
\includegraphics[5cm,14cm][17cm,2cm]{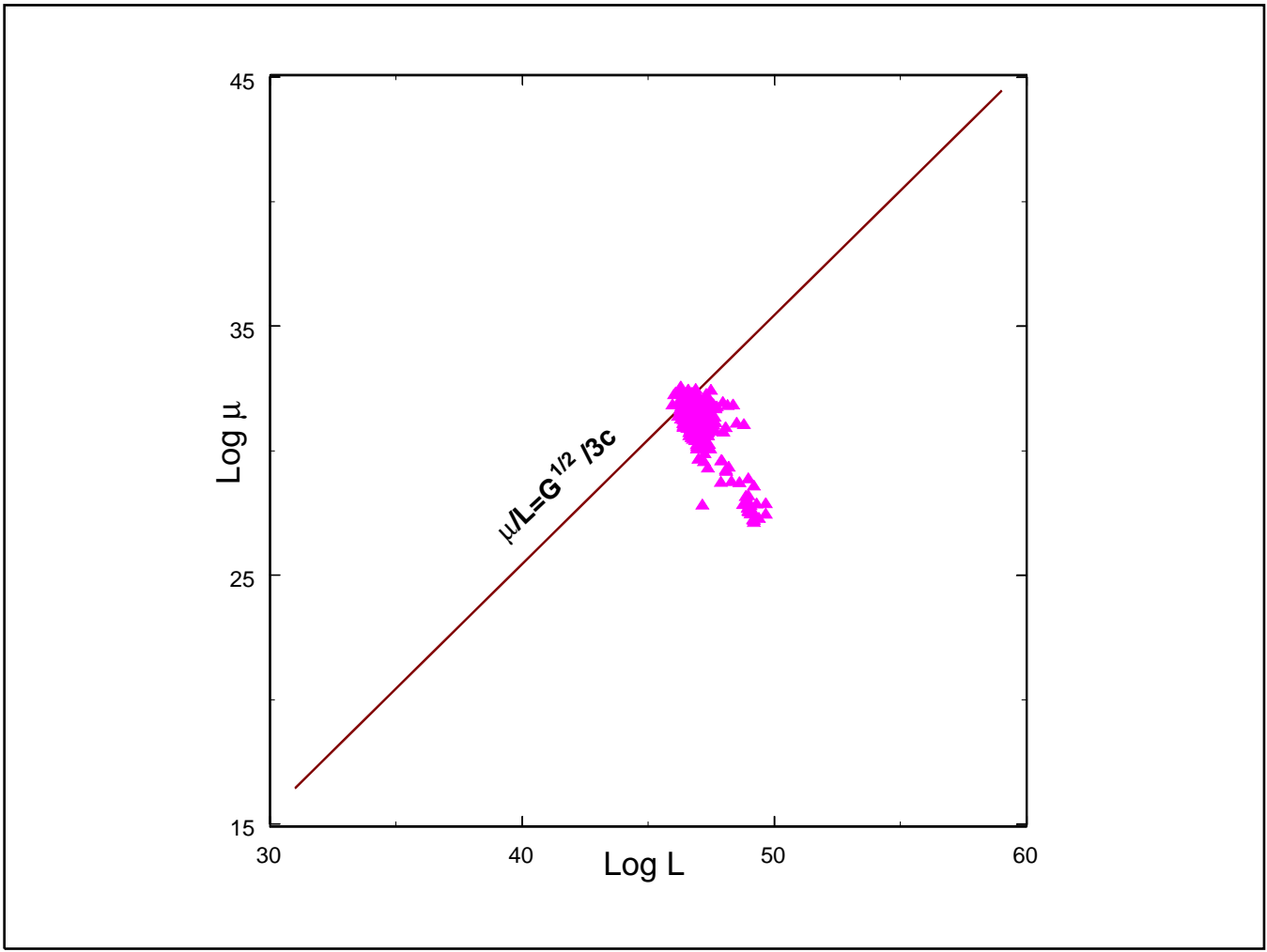}
\vspace{14cm} \caption{The estimated values of the magnetic
moments of pulsars \cite{18} vs. their angular momenta. Solid line
is Eq.({\ref{370}}). Axes are like in Fig.2.} \label{fig3}
\end{center}
\end{figure}

The gyromagnetic relations are measured  for three pulsars only
\cite {2}. These values are in a quite satisfactory agreement with
Eq.({\ref{370}})(Fig.2). For the majority of pulsars \cite{18},
there are estimations of magnetic fields obtained using a number
of model assumptions \cite{2}. It is impossible to consider these
data as the data of measurements, but nevertheless they also are
in some agreement with Eq.({\ref{370}}), (Fig.3).

For planets the situation is more difficult. First, inside planets
the substance forms not relativistic electron-nuclear plasma, but
nonrelativistic electron-ion plasma. It has different equation of
state \cite{10} leading to a more complex expression for the
stable mass of a planet core than the expression of
Eq.({\ref{230}}) for stars. Second, a noncharged layer at the
surface of the core can take a significant part of a planet's
volume and it is impossible to neglect a role of this stratum.
However, it can be seen from Fig.2 that the gyromagnetic relations
of planets are also in the quite satisfactory agreement with
Eg.({\ref{370}}). The detailed calculation for the Earth \cite{3}
gives for the magnetic moment  $4\cdot{10^{25}} Oe\cdot{cm^3}$,
which is almost exactly twice smaller than the measured value
$8.05\cdot{ 10^{25} Oe\cdot{cm^3}}$. Thus, it is possible to
assume that the basic component of the magnetic moment of planets
is induced by the same mechanism which is working in stars.

\clearpage


\clearpage



\begin{thebibliography}{18}

\bibitem {1}   Borra E.F. and Landstreet J.D. - The Astrophysical Journ, Suppl., 1980, v.42, 421-445.
\bibitem {2}   Beskin V.S.,Gurevich, A.V., Istomin Ya.N. - Physics of the Pulsar Magnetosphere, Cambridge University Press, 1993.
\bibitem {3}   Vasiliev B.V. - Nuovo Cimento B,1999,v.114,pp.291-300.
\bibitem {4}   Shiff L.I. and Barnhill M.V. - Phys.Rev.,1968,v.151,pp.1067-1071.
\bibitem {5}   Dressler A.I. a.o.  -Phys.Rev.,1968,v.168,pp.737-743.
\bibitem {6}   Riegel T.J. - Phys. Rev.B,1970,v. 2,pp.825-828.
\bibitem {7}   Kumar N. and Naddini R. -  Phys. Rev.D.,1973,v.7,pp.1067-1071.
\bibitem {8}   Leung M.C. et al. - Canad.Journ. of Phys.,1971,v.49,pp.2754-2767.
\bibitem {9}   Leung M.C. - Nuovo Cimento,1972,v.76,pp.825-929.
\bibitem {10}  Landau L.D. and Lifshits E.M. - Statistical Physics,1980, vol.1, 3rd edition,Oxford:Pergamon.
\bibitem {11}  Vasiliev B.V. and Luboshits V.L. -  Physics-Uspekhi,1994,v.37,pp.345-351.
\bibitem {12}  Kirzhnitz D.A. - JETP, 1960, v.38, pp.503-508.
\bibitem {13}  Abrikosov A.A. - JETP, 1960, v.39, pp.1797-1805.
\bibitem {14}  Leung Y.C. - Physics of Dense Matter, 1984, Science
\bibitem {15}  Heintz W.D. - Double stars,1978, Geoph. and Astroph.monographs, vol.15, D.Reidel Publ.Comp.
\bibitem {16}  Sirag S.-P. - Nature,1979,v.275,pp.535-538.
\bibitem {17}  Thorsett S.E. and Chakrabarty D. - E-preprint: astro-ph/9803260, 1998, 35pp.
\bibitem {18}  Taylor J.H., Manchester R.N., Lyne A.G., Camilo F., Catalog of 706 pulsars,
 1995, pulsar.prinston.edu

\end{thebibliography}
\end{document}